IAC-22-B3.7.5

# Categorisation of future applications for Augmented Reality in human lunar exploration

**Paul Topf Aguiar de Medeiros[a], Paul Njayou[b], Flavie A. A. S. D. T. Rometsch[c], Dr. Tommy Nilsson[d], Leonie Becker[e], Dr. Aidan Cowley[f]**

[a] *European Space Agency (ESA), European Astronaut Centre (EAC), Linder Höhe, 51147 Cologne, Germany*
hello@pauldemedeiros.nl
[b] *European Space Agency (ESA), European Astronaut Centre (EAC), Linder Höhe, 51147 Cologne, Germany*
paul.njayou@stud.ph-weingarten.de
[c] *European Space Agency (ESA), European Astronaut Centre (EAC), Linder Höhe, 51147 Cologne, Germany*
*flavie.rometsch@ext.esa.int*
[d] *European Space Agency (ESA), European Astronaut Centre (EAC), Linder Höhe, 51147 Cologne, Germany*
*tommy.nilsson@esa.int*
[e] *European Space Agency (ESA), European Astronaut Centre (EAC), Linder Höhe, 51147 Cologne, Germany*
leonie.becker1010@gmail.com
[f] *European Space Agency (ESA), European Astronaut Centre (EAC), Linder Höhe, 51147 Cologne, Germany*
aidan.cowley@esa.int

**Abstract**

The European Space Agency (ESA) has a clear mission to go forward to the Moon in preparation of human presence on Mars. One of the technologies looked at to increase safety and efficiency of astronauts in this context is Augmented Reality (AR). This technology allows digital visual information to be overlaid onto the user's environment through some type of display or projector. In recent years separate studies have been conducted to test the potential value of AR for astronauts by implementing a few functionalities on an AR display followed by testing in terrestrial analogue environments. One of the groups contributing to these investigations is Spaceship EAC (SSEAC). SSEAC is a group of interns and trainees at the European Astronaut Centre (EAC) focusing on emerging technologies for human space exploration.

This paper presents an outcome of SSEAC's activities related to AR for lunar extravehicular activities (EVAs), in which an approach similar to design thinking was used to explore, identify, and structure the opportunities offered by this technology. The resulting categorization of AR use cases can be used to identify new functionalities to test through prototyping and usability tests and can also be used to relate individual studies to each other to gain insight into the overall potential value AR has to offer to human lunar exploration.

The approach adopted in this paper is based on the Fuzzy Front End (FFE) model from the innovation management domain. Utilising a user-driven instead of technology-driven method resulted in findings that are relevant irrespective of the hardware and software implementation. Instead, the outcome is an overview of use cases in which some type of AR system could provide value by contributing to increased astronaut safety, efficiency and/or efficacy.

     An initial overview of AR functions for lunar EVAs was created based on existing literature. These were expanded on through a multidisciplinary brainstorm within SSEAC. A subsequent clustering activity resulted in a categorisation of potential AR applications.

The following categories were defined: EVA navigation, Scientific measurements and observations, Sample Collection, Maintenance, Repair, Overhaul (MRO) and Construction, Logistics and Inventory Management, Medical Procedures, Biomedical and System Status Monitoring, Collaboration and Support.

**Acronyms / abbreviations**

| | |
|---|---|
| AR | Augmented Reality |
| COTS | Commercial Off The Shelf |
| ESA | European Space Agency |
| EVA | Extravehicular Activity |
| FFE | Fuzzy Front end |
| SLS | Space Launch System |
| HUD | Heads Up Display |
| ISS | International Space Station |
| MRO | Maintenance, Repair, Overhaul |
| NASA | National Aeronautics and Space Administration |
| xEMU | eXploration Extravehicular Mobility Unit, |

## 1. Introduction

The international aerospace community is once again preparing for the exploration of the lunar surface by astronauts. Leading up to the anticipated crewed Artemis missions, scientists and engineers are working to define what lunar exploration will look like in the 21st century. Humanity has come a long way since the Apollo era, and one should expect higher standards of safety, increased science return and hopefully missions with a longer duration leading to the establishment of a sustainable human presence on the Moon. The new technological paradigm affects every single aspect of future missions, from the suits used during Extravehicular Activities (EVAs) to the communication infrastructure and the tools used for in-situ science and sample return.

    This paper presents the results of a project which aimed to create an overview and classification of potential use cases of Augmented Reality (AR) in the context of Lunar EVAs. Through a review of literature, a list of applications which have been investigated was made. Subsequently a guided brainstorm served to generate new ideas and concepts for novel use cases. Through a clustering activity, all the use cases were grouped together, and a classification was made to describe distinct application areas.

    The aim was not to make a fully comprehensive categorization, but rather to lay the groundwork for a user-centred design approach which can take these and other application areas into account in the design and development of the entire AR system. Secondarily, the overview made in this project can be helpful to others wishing to evaluate the potential benefits of AR for lunar EVAs across use cases. This more complete view of the benefits which could be derived from such a technology development could aid in decision-making regarding the allocation of funds for a lunar EVA AR interface.

    This paper is the result of an investigation into the potential of AR applications for Lunar exploration which was performed by interns at the European Astronaut Center (EAC) and more specifically within the Spaceship EAC group. This group consists of interns and trainees and aims to investigate low Technology Readiness Level technologies for space exploration.

*1.1 Lunar exploration context*
Although there have been fluctuations in the level of interest in and funding for human space exploration since the end of the Apollo program, there are indications that the current upwards trend will continue. There is international support for a strategy in which human exploration of the Moon will be used as a steppingstone towards human exploration of Mars [1]. This year NASA's Space Launch System (SLS) and the Orion spacecraft, a collaborative achievement between NASA and ESA, are scheduled to launch as part of the Artemis I mission. This inaugural uncrewed mission will prove the system's capability to bring humans into Lunar orbit. Meanwhile, an international collaboration of space agencies has started working on the next long-term human orbital outpost called 'Lunar Gateway', for the first time in history to be built in Lunar orbit. NASA's next-generation EVA spacesuit is also being developed with Lunar surface operations in mind [2]. The Human Landing System is the last piece of the puzzle which will allow astronauts to access the Lunar surface, and its development is being funded by NASA [3].

    Later phases of the Artemis program aim to establish longer-duration crewed lunar missions. ESA has also envisioned the establishment of an international lunar village, an outpost for long duration manned planetary missions. This would be an ideal platform not only for detailed science, but also to prepare for the first manned Mars missions [4].

    With the prospect of increasing human deep space exploration, the return of planetary EVAs and all the challenges related to long-term astronaut presence on lunar and planetary surfaces, we must consolidate efforts to develop optimized state-of-the-art technologies and tools to enable astronauts to work safely and efficiently.

*1.2 Augmented reality*
One such technology which has gained some interest in the context of EVAs is AR. Augmented reality involves the overlay of digital information onto the user's physical environment. There are three main types of AR technologies currently on the market [5]: Optical See-through AR consists of a transparent display which allows the user to see their physical environment behind digital projections. Video See-through AR, commonly used in Mobile Augmented Reality found on smartphones, consists of a display which shows a real-time video feed from a camera with overlayed digital information. Finally, Spatial AR does not make use of a display, but rather projects digital information directly onto the physical environment.

Augmented reality emerged several decades ago and has since then been developed in a multitude of technologies for various applications. Some of the earliest examples were found in military cockpits to aid pilots. Other use cases have been found in education, training, industry and more. The use of Augmented Reality for astronauts is also not a new concept. As far back as the 1980s and 90s, concepts were made for Heads up displays (HUDs) to be integrated in EVA suit helmets [6][7]. Practical experience has since been gained in microgravity through experimentation with both bespoke and Commercial Off the Shelf (COTS) AR interfaces on board the International Space Station (ISS)[8][9].

*1.3 AR for lunar exploration*
The integration of a HUD system has been documented as being one of the design goals for NASA's next generation EVA suit, called xEMU [10]. Although there have been some published tests with AR in the xEMU helmet [11], based on the lack of publicly available information it appears that this functionality is currently not on the critical design path for the system.

Numerous studies have been performed in which some specific functionality was implemented as a prototype on either bespoke or COTS hardware, to enable testing of AR functionalities in use cases analogous to astronaut operations in space [7] [12] [13] [11] [14] [15] [16] [17] [18] [19] [20] [21] [22] [23] [24] [25]. With some exceptions, the studies do not tend to adopt user-centred design processes, instead opting to work with available technology to demonstrate the benefits of AR in a specific use case.

In the setup of these studies, it is rarely mentioned why the hardware used for the study was chosen. If it is mentioned, it tends to be in the form of an evaluation of a few available options, comparing the suitability of these technologies to the specific use case intended for the study. There seems to be a knowledge gap concerning the wider context of potential applications for AR. This makes it difficult to select optimal technologies and system architectures for development, since one cannot predict the suitability of any given technology for all use cases if no overview of use cases exists.

The practical studies listed above choose a few highly specific use cases or applications, but do not tend to elaborate on how the choice for a specific use case was made, beyond establishing that they are relevant to the human space exploration context. This presents a limitation in the state of the art, since one must assume that a complex and presumably expensive system such as an AR interface rated for use inside an EVA suit, should be used for as broad a range of applications as is possible and useful.

Although individual studies have contributed significantly to showing applications of AR technology for human space exploration and the benefits which can be derived from them, there seems to be a need for a more comprehensive study of potential applications of this technology [26]. Such an overview would allow for a better understanding of the full benefits which can be derived from an AR system across use cases, which could form a stronger basis for the allocation of the necessary funding to develop such a system. Additionally, understanding potential use cases of AR irrespective of the technology used for implementation allows for a user-centred instead of a technology-driven design approach.

**2. Approach**
The aim of this project was to create an overview and classification of potential use cases of AR in the context of Lunar EVAs. The adopted approach finds similarities in the 'Fuzzy Front End' (FFE) phase of the product development process from the innovation management domain.

Defined as "the period between when an opportunity is first considered and when an idea is judged ready for development" [27], the FFE approach assumes that significant value can be derived from properly understanding the contexts, stakeholder needs and problem definitions of a new product before investing heavily into its development. This is reflected in the first half of the British Design Council's Double Diamond model for a structured design approach (Figure 1) [28], a widely utilized model in the Industrial Design Engineering industry. FFE also shares common attributes with the widely known 'Design Thinking' approach which emphasizes a human-centred, iterative approach including analysis and synthesis phases which employ, amongst other things, brainstorms, and clustering activities [29].

FFE aims to develop more optimized products by spending time to properly understand what is being developed and why. This should result in a higher return on investment and can prevent costly late-stage design changes which might incur significant delays in the delivery of a product or system [30]. Additionally, integrating relevant data in new ways during a well-structured FFE phase can lead to novel and innovative solutions. [31]

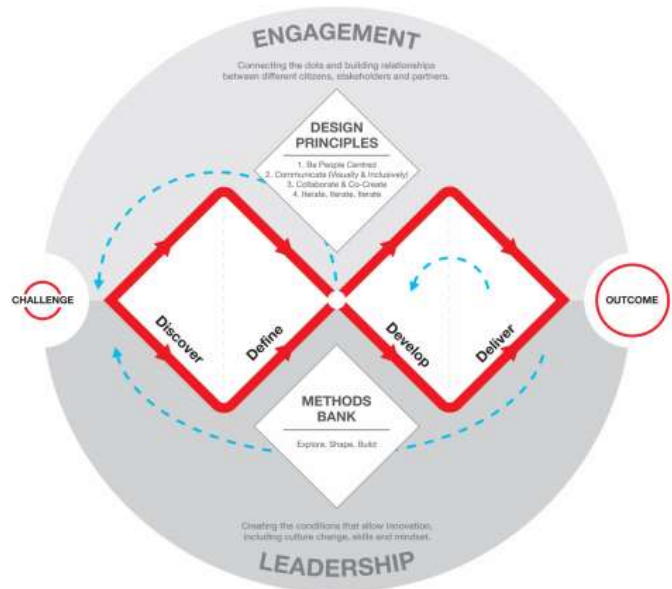

*Figure 1, the Double Design approach as described by the British Design Council* [28]

Characteristics of a well-structured FFE phase tend to be multi-disciplinary, collaborative, and iterative. The process should consist of multiple rounds of convergent and divergent activities and can include guided brainstorm sessions with experts, users and/or stakeholders. This allows a learning process to take place in which the problem is further defined, the user is better understood, and the context is further mapped out. Breuer et al. describe a classic FFE approach in which certain inputs are given to an expert brainstorm, which triggers a wide range of ideas (divergent) which are subsequently clustered (convergent) to form search areas. These search areas can then form the basis for further investigation, definition, ideation, and design (Figure 2) [32]

The classification generated in this project can be seen as analogous to the search areas in FFE, in that they do not specify a design or technology but rather represent insights into user needs and context factors such as science goals, and form demarcated areas which aid further ideation and concept development, breaking free from convention and existing assumptions about the applications of AR to develop user-centred solutions.

The approach to forming the classification also reflects processes commonly employed in FFE. Starting with contextual research, existing literature was studied to create an overview of applications which have previously been described and/or investigated. Subsequently, a guided brainstorm with a multi-disciplinary team of SSEAC interns and staff served to generate a large quantity of ideas for potential use cases. These were then clustered to create a categorization of AR use cases for lunar surface exploration. Finally, the categorization was tested against the applications described in literature to ensure they were representative of the existing body of work.

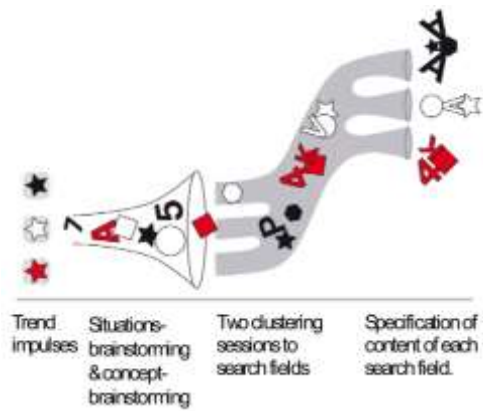

*Figure 2. The iterative divergent and convergent process as described by Breuer et al.* [32]

During the project it was decided to limit the scope to applications and use cases of AR during lunar EVAs. Although an even wider evaluation of applications for all elements of a human lunar exploration mission could be valuable, the more limited scope helped to gather useful insights within the limited timeframe of the project.

Publications related to among others NASA's IDEAS system, Holo-SEXTANT, SUITS program

were included in the review of existing literature. Due to the scope of the project, publications related to real-world experiments with AR in terrestrial industry and on the ISS such as ESA's MobiPV4Hololens were purposefully omitted. Inclusion of a wider selection of studies could be beneficial to find more potential applications, however the limitation of the scope was instrumental to complete the project within its limited timeframe.

The guided brainstorm was organized on August 3, 2020. Due to restrictions related to the COVID-19 pandemic, the brainstorm was organized remotely, and an online whiteboard tool was used in conjunction with video conference software. This allowed a group of interns, trainees, and staff from SSEAC with a wide variety of backgrounds from computer science to aerospace engineering and industrial design to join the session and contribute to the ideation of potential use cases of AR for human lunar exploration.

The first step in the brainstorm was not to directly talk about AR applications for lunar exploration. Instead, the 'principle of detour' [32] was applied and participants were asked to write down abstracted potential values offered by AR technologies regardless of their application area. Additionally, participants were asked to write down as many activities as they could think of that could possibly be a part of future human lunar exploration, without thinking about AR at all.

Subsequently, participants were asked to combine these two inputs and generate a large number of use cases. They were also instructed that not all use cases had to be linked to inputs which were defined in the previous step. To the contrary, the synthesis of use cases from insights should ideally trigger new ideas and insights, thereby leading to the identification of more use cases. The brainstorm lasted 2.5 hours, and the resulting use cases are described in section 3.

After the divergent phase, the seemingly random and chaotic collection of ideas needs to be ordered in some way. More than just an organization of ideas, the process of clustering also adds value to the creative process. By linking ideas together and choosing specific words to describe idea-spaces, new search areas are created which can form the basis for whole new concepts to be developed [33], indirectly triggered by the earlier discovery and definition steps.

The clustering activity was performed by two authors, in an iterative process that included feedback from other co-authors. The resulting classification can be found in the section 3.

Finally, the classification was tested against the applications found in existing literature. Through this process, it was realized that there was no category accurately representing the display of telemetry from various external sensors and that science operations outside of geological sampling had not been discussed during the brainstorm. To address this, a category was added to represent these use cases.

## 3. Results

16 publications were included in the review of applications mentioned and/or investigated by existing literature. Table 1 shows an overview of the applications per publication, worded as they are in the original text.

| Reference | AR applications which are investigated or suggested |
|---|---|
| Griffin, B. (1990)[7] | Map-type graphics for navigation, pre-recorded video instructions, remote live-streamed video from cameras, gauge readings for consumables |
| Hogson, E. et al. (2003) [12] | Life support and comfort control, communications, mission and task planning, localization and situational awareness, navigation, task execution |
| Di Capua, M. (2008) [13] | Life support and comfort control, mission and task planning, localization and situational awareness, navigation, task execution and human-robot interfaces |
| Stolen, M. et al. (2008) [11] | Monitor the status of their own and other's biometrics, monitor the status of their and other's spacesuit systems, monitor the status of robotic systems |
| Jacobs, S. et al. (2009) [14] | Navigation package, remaining consumables, crewmember health, suit status |
| Villorin, A. (2016) [15] | Procedure lists and task instructions, consumables status, camera tools, video communications, sensor telemetry views |
| Morrison, M. et al. (2017) [16] | Procedure checklists, navigational aids, display of biomedical data |
| Anandapadmanaban, E. et al. (2018) [17] | Traverse plans |
| Gibson, A. et al. (2018) [18] | Obstacle avoidance and wayfinding |
| Mitra, P. (2018) [19] | Cuff checklist, suit data display, camera control, communications, caution and warning system |
| Valencio D'souza, G. (2019) [20] | Maintenance task, navigation and rocks sample collection task |
| Fox, K. (2020) [21] | Task instructions |
| McHenry, N. et al. (2020) [22] | Visual display of suit vitals, telemetry, waypoints and checklist items |
| Radway, S. et al. (2020) [23] | Task instruction, sampling assistance, note taking, telemetry monitoring and display |
| Rometsch, F. (2020) [24] | Geological site inspection, data logging, photo documentation, taking site coordinates, verbal field notebook, waypoints, display of suit diagnostics |
| Miller, L. et al. (2021) [25] | Livestream of biometric values, procedure overview, reference resources to support activities with detailed information |

*Table 1: Applications described and investigated in existing literature.*

To generate a list which is more workable than the information in table 1, the list in table 2 was made, somewhat generalizing, and grouping specific applications together.

| Application | References |
|---|---|
| Navigation | [7][12][13][14], [16][17][18][20][22][24] |
| Procedure information | [7][12][13][15][16][19][20][21][22][23] |
| Camera live feed | [7][15][19] |
| Consumables monitoring | [7][11][14][15] |
| Life support control | [12][13] |
| Communications | [12][13][15][19] |
| Procedure planning | [12][13] |
| Situational awareness | [12][13] |
| Human-robot and Human-machine interfaces | [13] [11][15] |
| Biometrics monitoring | [11] [14][16] |
| Suit system status monitoring | [11][14][19][22][23] |
| Note taking and data logging | [23][24] |

*Table 2: Generalized overview of applications described and investigated in existing literature.*

As described in the approach section, a brainstorm was organized in which participants were asked to document ideas for potential values derived from AR irrespective of application type, and to document potential activities which might be a part of future human lunar exploration missions.

The following types of value which could be derived from a lunar AR system were identified:

For astronauts
- Reduce cognitive load
- More agency in accessing data
- Increase amount of information crew can access
- Enhance capabilities to control vessel in flight
- Easier crew to crew communication
- Free hands
- Increased situational awareness
- Ground can send information directly to crew's feed
- Enhanced communication between astronauts and ground
- Faster assembly / maintenance
- Decrease time needed to perform a task
- Live adaptable instructions
- Visual text-based communication messages
- Sharing target of attention
- Extend visual senses
- Ability to reconfigure the multipurpose interface
- Adaptable setting
- Integration into existing hardware

Programmatic value
- Enhanced PR content
- Lower risk for accidents
- More collaborative possibilities
- Increased general well-being of astronauts
- Avoid distractions for astronauts
- Increase astronauts' focus
- Less need for training
- Improved emergency response

The following lunar activities were described:
Gateway
- Communication, planning and preparation of day-to-day tasks
- Crop cultivation
- Hardware troubleshooting
- Tele-medicine
- Payload deployment
- Retrieving regolith samples
- Hardware status observations
- Construction of infrastructure
- Post- and pre- EVA activities
- Performing experiments
- Leak detection
- In-Situ medical care
- Post- and pre- flight activities
- Spare part manufacture
- Payload maintenance
- Resting/ sleeping
- Cargo and stowage logistics
- Payload upgrading

Human Landing System
- Dust mitigation in habitat
- Collaboration between Gateway and lunar surface

- Terrain awareness
- Live flight data
- Hardware troubleshooting
- Retrieving/handling regolith samples
- Tele-medicine
- Preparing samples for return to Earth
- System integrity checks
- Leak detection
- Communication, planning and preparation of day-to-day tasks
- Resting/sleeping
- Post- and pre- EVA activities
- Proof of concepts for fuel and oxygen storage and transportation
- In-situ medical care
- Synthetic landing site markers

Lunar surface
- Harvesting lunar volatiles
- Terrain awareness
- In-situ analysis of geological samples
- Crop cultivation
- Tele-geology
- Exploration of Permanently Shadowed Regions
- Retrieving regolith samples

- Hardware troubleshooting
- Dust mitigation on equipment
- Tele-medicine
- Traverse over rough terrain
- proof-of-concepts for fuel and oxygen storage and transportation
- Co-bot operations
- Performing experiments
- Mapping and characterization of macro geological features
- Construction of infrastructure
- Leak detection
- Construction of roads or landing pads
- Live checklists
- Communication, planning and preparation of day-to-day tasks
- Spare part manufacture
- Construction of infrastructure
- In-situ medical care

Subsequently, participants were asked to write down as many use cases of AR for lunar exploration as they could come up with. Each use case should have a title, and one or two sentences detailing the function and added value of AR in this use case (Table 3).

| | Use Case Title | Function | Value |
|---|---|---|---|
| 1 | Rover / instrument maintenance | Display procedures, schematics to do maintenance work on an instrument | Less training required as procedures are automatic and updated accordingly, easy to follow and highlights and displays overlays on the |
| 2 | Construction of roads / landing pads | Helps astronauts in selecting areas to construct basic infrastructure and helps them in finding level ground to build on. | Support for construction tasks that would require additional hardware, integrated into a HUD. |
| 3 | Instructions Overlay | overlay visual assembly or maintenance cues (highlight next screw holes, insertion path/orientation of parts etc.) | Faster assembly / maintenance, less training required, fewer errors. |
| 4 | Sample selection HUD | HUD provides overlay of information from an IR camera to provide more information about potential sample composition | Increased science return from samples more efficient use of astronaut time |
| 5 | Communication between astronauts during EVA | HUD allows astronauts to communicate by highlighting physical objects, and by transferring data from one to another (e.g., location, health monitoring). | Reduces the likelihood of misunderstandings, increases the ability of astronauts to assist each other (e.g., rescue), makes communication more effective, decreases the amount of verbal communication needed. |
| 6 | Sample retrieval | Display the location of a sample and protocols to follow for retrieval | Minimize sample retrieval time |
| 7 | Classic flight / landing HUD | Will display flight data, landing data and environmental data on a classic HUD allowing astronauts to observe the Lunar environment during critical phases. | less accidents, better situational awareness |
| 8 | Non-vocal one-way communication | Messages by ground control or Gateway can be sent to the astronaut's HUD and displayed there. | no need for vocal communication |
| 9 | Medical information in HUD | Displaying personal vitals and vitals of crew members. Basic vitals (e.g., blood pressure, heart rate, O2sat). Can also display energy expenditure and give warnings if overexerting oneself. | Reduces the need to request medical information. Can increase safety, increase emergency response |
| 10 | Checklists in HUD | Checklists of items (i.e., deployment of stuff, or procedures). Collaborative checklists could possibly be synchronized in real time. | No need for an additional device for checklists |
| 11 | Construction enhancer | Simulate beams and loads and payloads to calculate the optimal structure or deployment | |
| 12 | Mission markers | Visual representation of items to be interacted with | Good overview of where to go for the next objective |
| 13 | Remote support during medical operations | Enables an expert on the ground (i.e., medical doctor) to provide relevant visual information to an astronaut performing a minor surgery. This information can be: checklists in text format, pre-recorded visual instructions, virtual pointer/highlighting to guide astronaut, live video feed from instructor. | Reduces the amount of training needed, increases the odds of success of surgery, increases the flexibility in terms of performable operations (instructor can adapt to exact situation) |
| 14 | Telepresence of expert / instructor | Overlay of video-feed of expert or instructor enabling additional communication channels (gestures, demonstration of movements etc.) | Higher quality communication, easier interaction with instructor or expert |
| 15 | EVA mini map | Display current position around ISS, or on lunar and/or planetary surface relative to base camp (including surface features etc. from satellite imagery) as well as teammate [Gä1] 's positions. | Increased situational or locational awareness of self and crew. This is good for safety, efficiency, and cooperation. |

*Table 3: Use cases resulting from the brainstorm*

After the brainstorm, the resulting use cases were clustered in a collaborative and iterative process amongst the co-authors of this publication. The following classification (Table 4.) was deemed to be representative of all use cases, while maintaining sufficient differentiation between each class. It should be noted that each class of use cases can contain multiple specific use cases and each use case can involve a combination of AR applications (e.g., waypoints, procedure list) and UI elements (e.g., video feed, overlaid data on the physical terrain). Table 4, 'related use cases from literature' only refers to use cases found in literature listed in Table 1

| Use case classification | Description | Related use cases from literature |
|---|---|---|
| **EVA navigation** | Navigation on the surface with or without vehicle. Positioning, situational awareness and interpretation of terrain features. | Navigation, Procedure planning, Situational awareness, Human-Robot and human-machine interfaces |
| **Scientific measurements and observations** | Observation and interpretation of data from science instruments, control of science instruments, annotation and tagging of data. | Camera live feed |
| **Sample collection** | Sample collection process, sample and site documentation and data logging. | Procedure information, procedure planning, Camera live feed |
| **MRO and construction** | Maintenance, Repair and Overhaul (MRO) and construction procedures, instructions, annotation, simulation, compliance testing and data logging. | Procedure information, procedure planning, Human-Robot and human-machine interfaces |
| **Logistics and inventory management** | Inventory tracking, equipment and consumables management, process and storage optimization. | |
| **Medical procedures** | Diagnostic, emergency, and scientific procedures. | Procedure information, procedure planning, Huma-Robot and human-machine interfaces |
| **Biomedical and system status monitoring** | Monitoring of crew member's vitals and critical system telemetry. | Consumables monitoring, Life support control, Human-machine interfaces, biometric monitoring, suit system status monitoring. |
| **Collaboration and support** | Collaboration between crew members, crew and ground, EVA crew and crew inside a habitat, lunar surface crew and Gateway crew or crew and (semi)-autonomous robotic systems. | Camera live feed, Communications, Human-robot and Human-machine interfaces |

*Table 4, classification of use cases of a lunar EVA AR*

## 4. Discussion

The results of this project encompass a wide variety of applications, and the classification should be useful in the generation of new concepts and the development of a user-centred system design.

Although efforts were made to include a wide variety of activities and use cases, the overview of use cases cannot be seen as comprehensive, even within the limited scope of lunar EVAs. This is evidenced by the fact that a significant group of activities was not found during the brainstorm and was instead added later, which indicates that there are likely to be other use cases which have not been found during this project. Ostensibly, making a complete overview of activities might not be possible until the actual mission profiles have been decided on. Until that time, one can however assume a certain value to be inherent in insights which aim to be diverse if not complete.

A certain transition is evident between the 'applications' of technology-driven design developments and evaluations - which constitute most of the existing literature - and the 'use cases' which are more relevant for the user-centred approach. The difference can be described as applications representing technical functions (i.e., placing waypoints, displaying a list of procedures, controlling the Life Support System, see 'Table 1') whereas use cases represent activities with more clear stakeholders, contexts and goals (i.e., 'guiding non-geologists during geological inspection tasks' [34]). The latter feeds directly into user-centred concept development and could allow designs to let go of conventions informed by the paradigm of outdated technologies. Any realistic system should however keep in mind the proven processes and designs which have been in use for decades. Future designs should incorporate these to benefit from their reliability and compatibility with existing systems.

Although a user-centred approach can lead to novel and optimized designs, one could argue that technical limitations should be given as much importance as design considerations as user needs. Especially for a technology which should work inside an EVA suit in use, extreme technical challenges need to be overcome to create a functioning system. For example, the electronics must be safe to use in the oxygen-rich environment inside a suit, integration of multiple systems such as GPS and IoT networks can rapidly increase complexity and cost, and redundancy must be built into systems which are critical for mission success and astronaut safety. All this considered, the technology-driven approach does not guarantee that these limitations are considered, since many studies are based on terrestrial COTS systems and would not fulfil these requirements. And a user-centred approach would include considerations for technical limitations in the design embodiment and detailing phases, as represented for example in the iterative 'develop and deliver' diamond shown in Figure 1.

This project has proven that there are relevant methodologies from the innovation management domain that could be applied to the development of complex systems for human space exploration. Future studies could potentially identify more opportunities for the development of user-centred systems for astronauts when applying methodologies from the innovation management and design engineering domains, as also evidenced by Rometsch et al. [35].

The main subject of this project was the classification of potential AR use cases for human lunar exploration. Although the outcome should be useful in its current form, one can imagine an even more comprehensive classification process which would not limit the scope to EVAs but to all activities related to human lunar and planetary exploration.

Furthermore, the approach which was used to create the classification could be formalized further, ensuring that the resulting categorization is comprehensive and individual classes are sufficiently differentiated from each other. An example of an excellent formalized classification of AR use cases was performed by Röltgen and Dumitrescu and could serve as an inspiration for further work in the subject area of this publication [36].

By focusing specifically on visual AR systems, the potential value of multi-modal AR systems might have been overlooked. Multi-modal AR systems use a mix of stimuli to provide data to the user instead of solely using visual displays. For example, Gibson et al. studied the use of haptic feedback in astronaut boots for obstacle avoidance [18]. Although challenging, it is likely worthwhile to include multi-modal interfaces as a consideration in the further development and evaluation of AR systems for lunar exploration.

## 5. Conclusion

This project has fulfilled its aim of generating a classification of potential use cases of AR for human lunar surface exploration. Although the scope had to be narrowed down to AR for EVAs, the hope is that future work can identify use cases for every potential context of use for an astronaut AR system . A more formalized process for classification might yield results which are more comprehensive with more precisely defined categories. However, it is expected that the results from this project already in their current form can help to evaluate potential AR technologies, support concept development of novel AR functions and provide a framework to bring together results from individual studies and start to form a picture of the full potential value which might be gained from the development of an AR system for human space exploration.

The following categories were defined: EVA navigation, Scientific measurements and observations, Sample Collection, Maintenance, Repair, Overhaul (MRO) and Construction, Logistics and Inventory Management, Medical Procedures, Biomedical and System Status Monitoring, Collaboration and Support.